\documentclass[9pt,nofootinbib,twocolumn]{revtex4}
\usepackage{amsmath,amsfonts,amssymb} 
\usepackage[pdftex]{graphicx} 
\usepackage{setspace}
\usepackage[margin=0.6in, paperwidth=9in, paperheight=11in]{geometry}
\usepackage{bbm}
\begin{document}

\begin{widetext}
\begin{center}
\textbf{Transverse optical and atomic pattern formation}
\vspace{5mm}

{Bonnie L. Schmittberger$^1$ and Daniel J. Gauthier$^{1,2}$}

\textit{$^1$Duke University, Department of Physics and Fitzpatrick Institute for Photonics, Box 90305, Durham, North Carolina 27708, USA}\\
\textit{$^2$The Ohio State University, Department of Physics, 191 West Woodruff Ave., Columbus, OH 43210 USA}
\end{center}
\vspace{5mm}

The study of transverse optical pattern formation has been studied extensively in nonlinear optics, with a recent experimental interest in studying the phenomenon using cold atoms, which can undergo real-space self-organization. Here, we describe our experimental observation of pattern formation in cold atoms, which occurs using less than $1~\mu\text{W}$ of applied power. We show that the optical patterns and the self-organized atomic structures undergo continuous symmetry-breaking, which is characteristic of non-equilibrium phenomena in a multimode system. To theoretically describe pattern formation in cold atoms, we present a self-consistent model that allows for tight atomic bunching in the applied optical lattice. We derive the nonlinear refractive index of a gas of multi-level atoms in an optical lattice, and we derive the threshold conditions under which pattern formation occurs. We show that, by using small detunings and sub-Doppler temperatures, one achieves two orders of magnitude reduced intensity thresholds for pattern formation compared to warm atoms.
\end{widetext}

\vspace{60mm}


Spontaneous pattern formation is a phenomenon that occurs in nearly all branches of science and has led to new insights into nonlinear dynamics on multiple scales of nature. In nonlinear optics and atomic physics, transverse optical pattern formation has been studied for more than three decades and has led to advances in all-optical switching and a better understanding of absolute instabilities~\cite{DawesReview}.

Transverse optical pattern formation describes the spontaneous formation of new optical fields via nonlinear optical wave-mixing processes. When optical fields counterpropagate through a nonlinear optical material, such as an atomic vapor, they induce a nonlinear refractive index in the atoms. Above a threshold nonlinear refractive index, an instability generates new optical fields, which are called optical patterns for their characteristic multimode, petal-like structure. Until recently~\cite{GreenbergOptExp,SchilkeExpTransverse,Labeyrie}, transverse optical pattern formation has only been studied using warm atomic vapors~\cite{Grynberg1988321,Firth90,Dawes29042005}. However, new physics is accessible when studying pattern formation in cold atoms.

When optical fields interact with cold atoms, they impart a dipole force that can influence the center-of-mass degrees of freedom of the atoms such that the atoms spatially bunch into real-space structures. Therefore, by studying the formation of optical patterns in cold atoms, one can also study the spontaneous formation of new atomic structures, also known as atomic self-organization. Self-organization has been studied extensively in single-mode cavities~\cite{PhysRevLett.91.203001,BaumannSelfOrganization}. However, by studying pattern formation in cold atoms, one gains access to a multimode geometry, in which one can study continuous symmetry-breaking and phase transitions that are inaccessible in a single-mode system~\cite{GopLevGoldNatPhys}.

In this paper, we present our experimental observation of pattern formation in sub-Doppler-cooled atoms, which was first reported in Ref.~\cite{GreenbergOptExp}. We show that we observe patterns at record-low powers, and we analyze the temporal dynamics of the power in the generated fields. We find that the optical patterns and the corresponding self-organized atomic structures~\cite{SchmittbergerMultimode} undergo continuous symmetry-breaking, which characteristic of non-equilibrium phenomena in a multimode geometry.

In addition, we present a theoretical description of pattern formation in cold atoms, which incorporates atomic self-organization into ``atomic patterns.'' Existing theoretical descriptions of real-space pattern formation in cold atoms are restricted to the regime of weak atomic bunching~\cite{Muradyan}, where the dipole potential energy of the applied lattice is comparable to the thermal energy of the atoms, or do not account for the possibility of longitudinal atomic pattern formation~\cite{GreenbergEPL, Labeyrie, PhysRevLett.112.043901}. The theoretical model presented in this paper is unique in that it is self-consistent and allows for tight atomic bunching, where one can achieve enhanced light-atom interaction strengths~\cite{PhysRevA.90.013813}. In addition, while other models for pattern formation describe self-organization in momentum space~\cite{Labeyrie}, our model explicitly considers the self-organization of atoms into three-dimensional real-space structures, which is of interest for studying non-equilibrium phenomena in cold atoms.

We use our model to derive the threshold conditions above which patterns form, and we provide an analysis of optimizing the nonlinear refractive index to achieve pattern formation at ultra-low light levels. We show that real-space self-organization of atoms gives rise to an enhanced nonlinear refractive index, which reduces the threshold powers required for pattern formation. We find that, by using small detunings and far-sub-Doppler temperatures, one can achieve more than an order of magnitude reduction in the intensity required to observe pattern formation \textit{cf.} warm atoms~\cite{Dawes29042005} and cold-atom experiments that work closer to the Doppler temperature~\cite{Labeyrie}.

The outline of this paper is as follows. We first briefly describe our experimental observation of pattern formation in cold atoms. We then go on to describe theoretically the interaction of counterpropagating optical fields with multi-level atoms. We analyze the nonlinear refractive index $n_{\text{NL}}$ for atoms in a one-dimensional (1D) optical lattice in the regime of strong light-atom interactions with multi-level atoms, and we discuss how to optimize $n_{\text{NL}}$ to reach the threshold for pattern formation at ultra-low intensities. We then extend this 1D model to two spatial dimensions and perform a stability analysis to derive the threshold condition under which pattern formation occurs. Finally, we compare these predictions to our experimental data.

\section{Overview of the experiment}
\begin{figure}
\begin{center}
\includegraphics[scale=0.27]{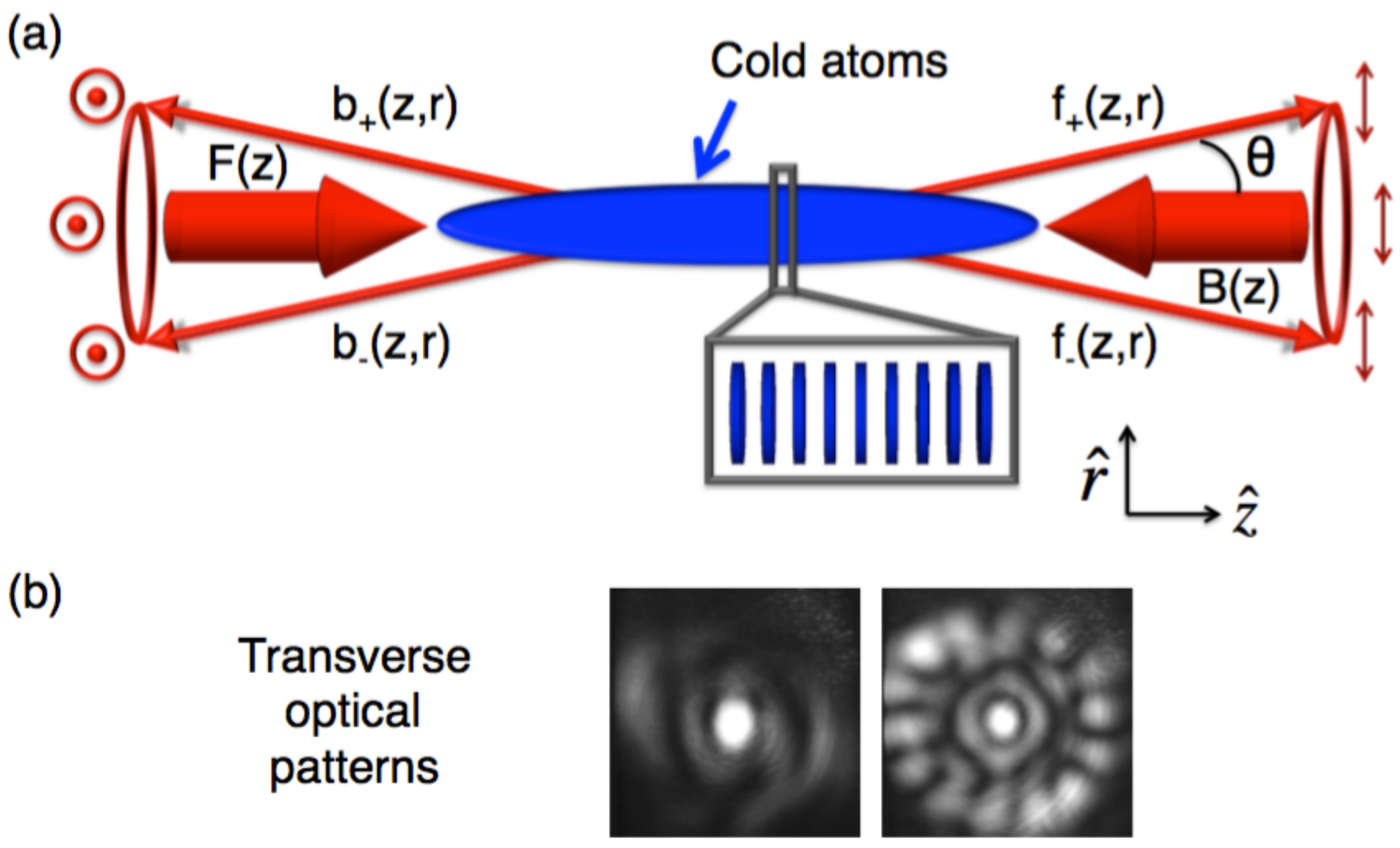}
\vspace{4mm}
\caption{\textbf{Experimental setup.} (a) We apply counterpropagating optical fields to a cloud of atoms of length $L=3~\text{cm}$ and width $w\sim400~\mu\text{m}$. The 1/e$^2$ beam waist of the pump fields is $\sim410\text{ }\mu\text{m}$. The arrows/circles denote the lin$\perp$lin electric field polarizations. The fields generated in the wave-mixing process propagate along a cone of half-angle $\theta$. We denote the fields propagating along or nearly along the ``forward'' $+\hat{z}$-direction as $F(z)$, $f_+(z,r)$, and $f_-(z,r)$, and those along the ``backward'' $-\hat{z}$-direction as $B(z)$, $b_+(z,r)$, and $b_-(z,r)$. (b) Examples of two- and fourteen-spot optical patterns, where the central spot is bleedthrough pump light, and the small ring closely surrounding the central pump spot arises due to a beam reshaping effect.}
\label{patterntheorypapersetupfig}
\end{center}
\end{figure}

In the experiment, we cool and trap a cloud of $^{87}$Rb atoms in an anisotropic magneto-optical trap (MOT) to an initial sub-Doppler temperature of $T_{\text{init}}\simeq30~\mu\text{K}$~\cite{Greenberg:07}, where the Doppler temperature is $T_D=146~\mu\text{K}$. We then apply counterpropagating optical fields in a lin$\perp$lin polarization configuration with electric field $\vec{E}=\vec{E}_0(z)e^{-i\omega t}+\text{c.c.}$ along $\pm\hat{z}$, as shown in Fig.~\ref{patterntheorypapersetupfig}(a), where
\begin{equation}
\overrightarrow{E}_0(z)=F(z)e^{ikz}\hat{x}+e^{i\phi}B(z)e^{-ikz}\hat{y},
\label{Efieldch7}
\end{equation}
$k$ is the wavevector in vacuum, and $\phi$ is an arbitrary relative phase between the fields. We typically use optical fields of frequency $\omega$ with a detuning $\Delta=\omega-\omega_{eg}\simeq-3\text{ to }-10\Gamma$ from the resonant transition frequency $\omega_{eg}$ of the $5^2S_{1/2}(F=2)\rightarrow5^2P_{3/2}(F^\prime=3)$ transition of $^{87}$Rb, where $\Gamma$ is the natural linewidth of the transition. Upon applying this 1D optical lattice, the atoms undergo Sisyphus cooling to a final temperature along $\pm\hat{z}$ of $T_z\simeq2-3~\mu\text{K}$~\cite{GreenbergOptExp}, while below threshold, the temperature along the radial direction remains $T_r\simeq30~\mu\text{K}$. After cooling, the thermal energy of the atoms is typically much less than the dipole potential of the applied 1D lattice, and the atoms spatially bunch into the potential minima and form pancake-like structures, as depicted in Fig.~\ref{patterntheorypapersetupfig}(a).

By using relatively small detunings and employing Sisyphus cooling, atoms tightly bunch at the locations of pure $\hat{\sigma}^\pm$-polarization, which gives rise to a bunching-induced enhancement of the nonlinear refractive index~\cite{PhysRevA.90.013813}. Above a threshold nonlinear refractive index, a transverse instability generates new nonlinear wave-mixing processes, which generate new optical fields (optical patterns) and atomic structures (atomic patterns).

\begin{figure}
\begin{center}
\includegraphics[scale=0.2]{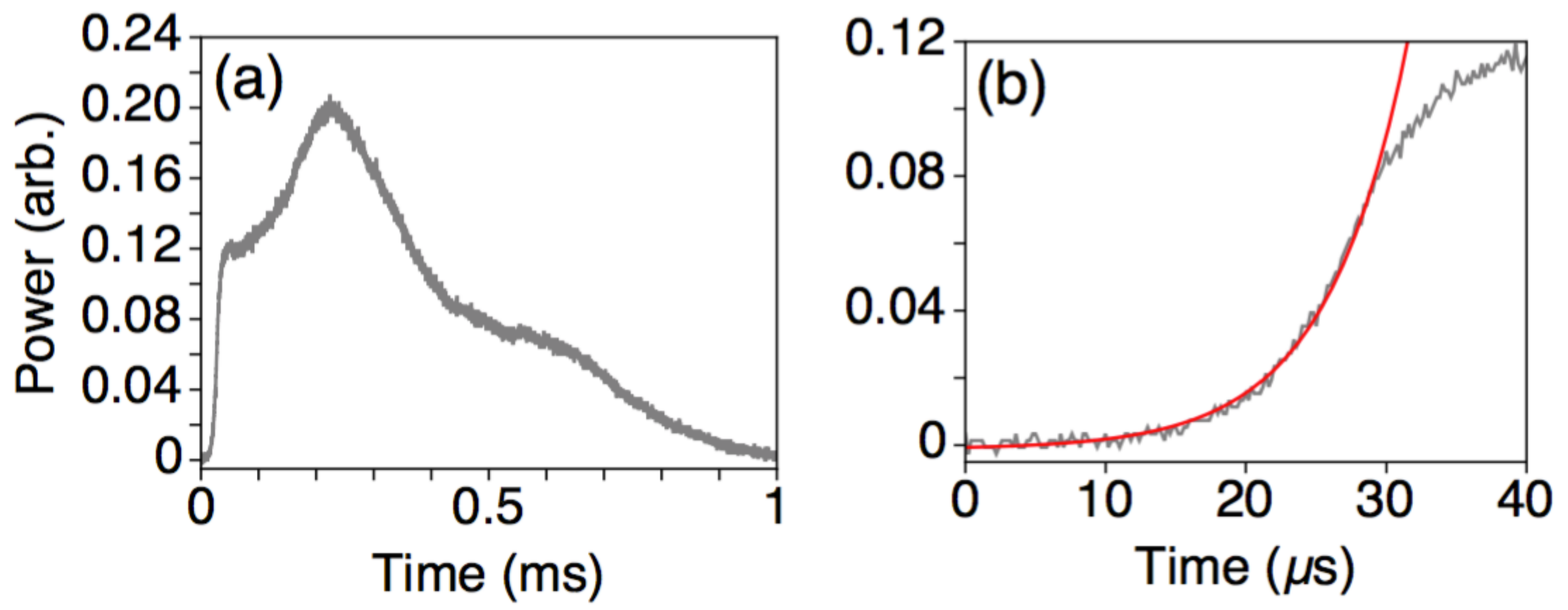}
\vspace{4mm}
\caption{\textbf{Exponential power increase.} (a) Power in the generated fields as a function of time for $I_p=12\text{ mW/cm}^2$ and $\Delta=2\pi\times(-24)\text{ MHz}$. (b) Zooming in on the formation of the patterns, the exponential fit (red, solid curve) to a function of the form $a\text{exp}[(t-t_\delta)/\tau]+b$ has an exponential constant $\tau=5.8\text{ }\mu\text{s}$.}
\label{exponentialbuildup}
\end{center}
\end{figure}

Once the patterns begin to form, there is an exponential buildup of the power in the optical patterns, as shown in Fig.~\ref{exponentialbuildup}. Figure~\ref{exponentialbuildup}(a) shows the typical temporal dynamics of the power in the optical patterns, where the pump beams are turned on at time $t_\text{on}=0$. Figure~\ref{exponentialbuildup}(b) provides a closer look at the behavior of the generated fields close to time $t_\text{on}$. Once a finite field begins to form, there is an exponential increase in the power generated in the wave-mixing process, which is the expected behavior for such wave-mixing instabilities~\cite{Yariv:77}. For typical experiments, we load the MOT for $97\text{ ms}$ and then shut the cooling and trapping beams off. We then turn on the pump beams for 3 ms to perform experiments, where the patterns persist for $1-2.4~\text{ms}$.

We observe the minimum threshold at a single-field intensity of $I_p=0.8I_{\text{sat}}$ or a power of ${\sim}400~\text{nW}$ at $\Delta=-4\Gamma$, where $I_{\text{sat}}=1.3~\text{mW/cm}^2$ is the resonant saturation intensity. The generated optical fields propagate along a cone of half-angle $\theta\simeq3-10~\text{mrad}$, which conserves momentum in the wave-mixing process~\cite{Chiao}. We observe frequency-degenerate pattern formation, so that all the optical fields generated have the same output frequency $\omega$ as the applied pump fields, and they have a polarization that matches the nearly counterpropagating pump field, as shown in Fig.~\ref{patterntheorypapersetupfig}(a). When imaged in the transverse plane, we observe optical patterns that have between two and fourteen spots, as shown in Fig.~\ref{patterntheorypapersetupfig}(b).

\begin{figure}
\begin{center}
\includegraphics[scale=0.18]{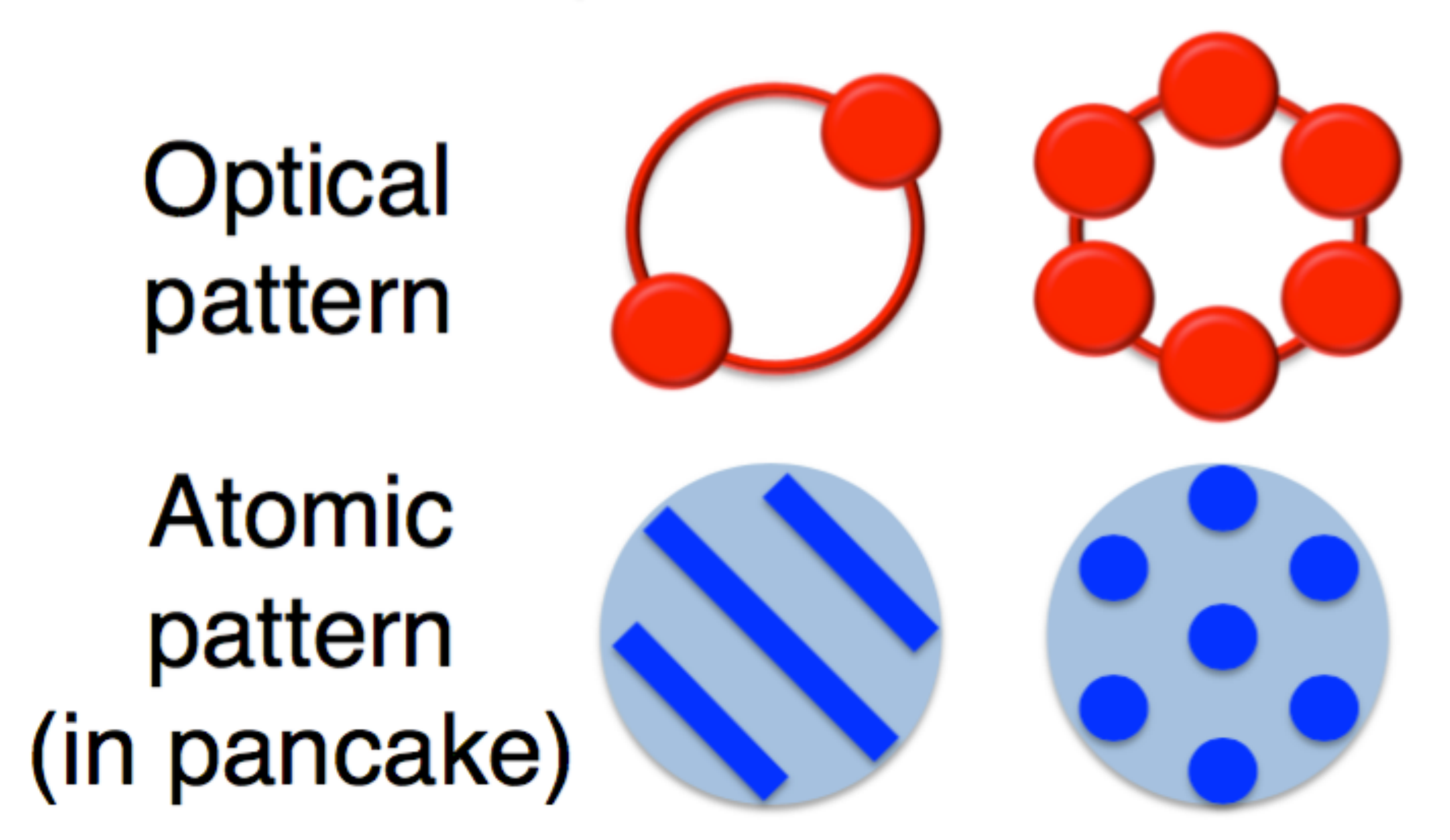}
\vspace{4mm}
\caption{\textbf{Simulations of self-organized patterns.} Two- (six-) spot optical patterns will generate striped (hexagonal) atomic patterns within each pancake.}
\label{patternsims}
\end{center}
\end{figure}

These optical patterns are correlated with associated atomic patterns in the atoms, which arise due to atomic bunching in the dipole potential wells generated by the interference among the pattern-forming optical fields and the applied pump fields. These atomic patterns, or self-organized atomic structures, occur within each pancake described in Fig.~\ref{patterntheorypapersetupfig}(a). For example, a two-spot optical pattern will create a striped-like interference pattern within each pancake, which in turn gives rise to striped dipole potential wells into which the atoms can bunch, as depicted in Fig.~\ref{patternsims}. Similarly, higher-order optical patterns will generate more complicated atomic patterns.

In Ref.~\cite{SchmittbergerMultimode}, we provide a direct observation of real-space atomic self-organization due to pattern formation in cold atoms. We use parametric resonance techniques to verify that the atoms bunch into the self-generated dipole potentials simulated in Fig.~\ref{patternsims}. We also find that, above threshold for pattern formation, the atoms also undergo spontaneous three-dimensional (3D) Sisyphus cooling due to the interaction among the generated and applied fields with the atoms. We use Bragg scattering techniques to show that the temperature of atoms along the radial direction cools to $T_r\simeq2-3~\mu\text{K}$ above threshold for pattern formation. Thus, in this paper, we do not account for Sisyphus cooling, which gives rise to a non-equilibrium gas~\cite{PhysRevA.69.013410}. Instead, we approximate the momentum distribution of the atoms using a simplified Maxwell-Boltzmann distribution characterized by the temperature of the atoms after they have undergone Sisyphus cooling ($T\simeq2-3~\mu\text{K}$).

\begin{figure}
\begin{center}
 \includegraphics[scale=0.18]{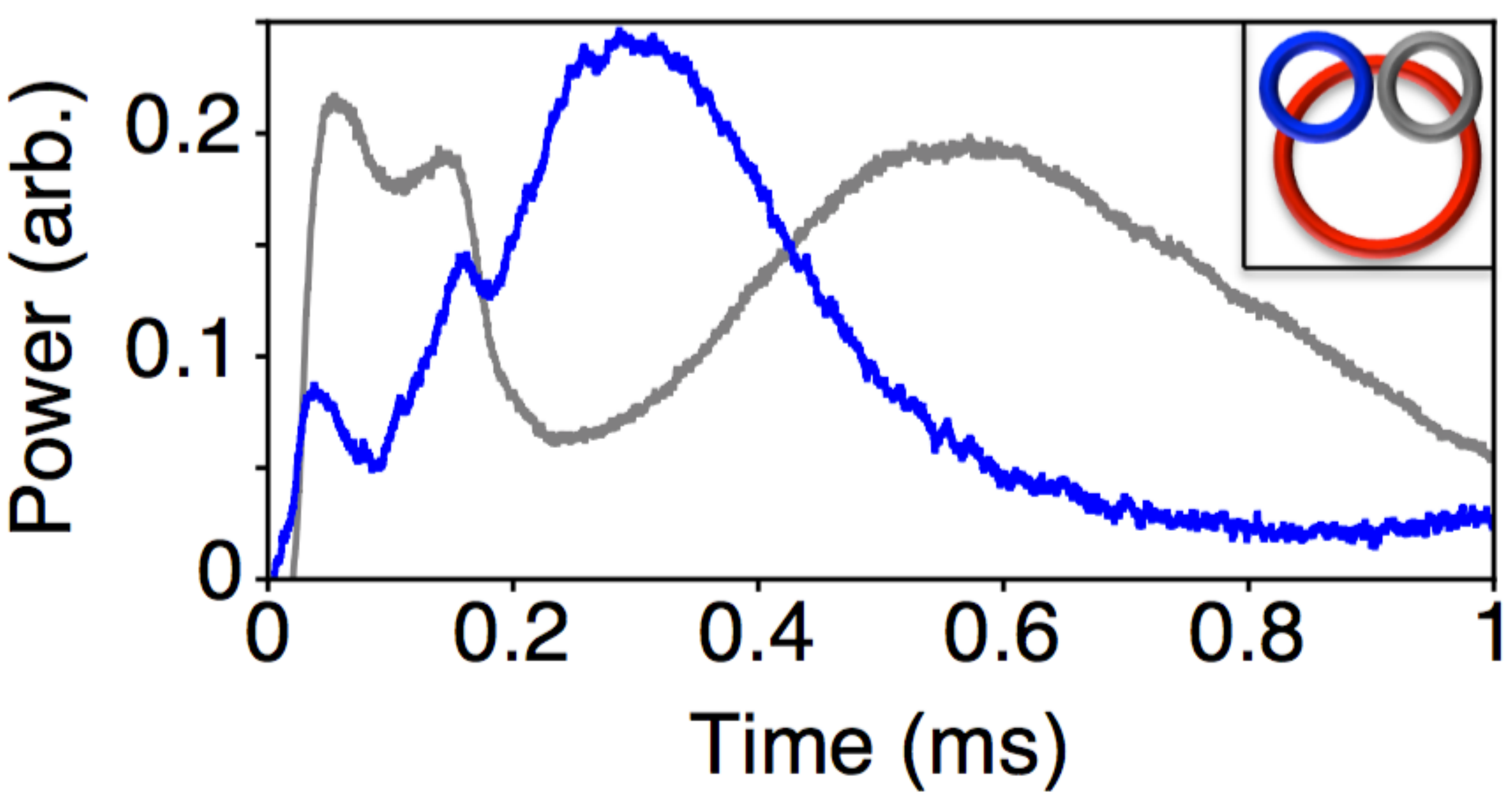}
 \caption{\textbf{Pattern rotation.} The power of the generated light within two distinct spatial locations on the cone of emission, as depicted in the inset, as a function of time. The pump beams are turned on at $t=0$.}
 \label{contsymmbreak}
  \end{center}
 \end{figure}

We also observe that the optical/atomic patterns are not necessarily stationary, and they can rotate/reorganize during a single experiment. To measure this pattern rotation, we split the path of the generated fields and place apertures that select two distinct emission regions, as depicted in the inset of Fig.~\ref{contsymmbreak}. The amplitudes of the generated field power through each aperture are shown together in Fig.~\ref{contsymmbreak} as a function of time. The observed anti-correlation between these modes (correlation factor = -0.2) indicates that the generated fields are rotating between these spatial modes. The fastest timescale for fluctuations between modes is ${\sim}50\text{ }\mu\text{s}$, which is the order of the time it takes for an atom to move a distance $d_p=\lambda^\prime/4\sim 195~\text{nm}$ and thus contribute to exciting a different optical/atomic pattern. This type of continuous symmetry-breaking is a hallmark of multimode non-equilibrium phenomena like pattern formation.
 
We observe optical/atomic pattern formation at record-low powers because we use sub-Doppler temperatures and small detunings. This experimental regime facilitates enhanced light-atom interactions because the atoms tightly bunch in regions of pure $\hat{\sigma}^\pm$ polarizations, where the atoms interact strongly with the optical fields. To theoretically describe pattern formation in this regime, we develop a new model that allows for tight atomic bunching in the applied lattice, as existing models are restricted to homogeneous~\cite{Firth90} or weakly bunched atoms~\cite{Muradyan}. We also require a model that accounts for real-space self-organization of the atoms during pattern formation in both the transverse and longitudinal dimensions. In the remainder of this paper, we present this theoretical model, which self-consistently describes optical/atomic pattern formation in both the weak- and tight-bunching regimes.

\section{Applied 1D optical lattice}
To describe theoretically the light-atom interaction that gives rise to pattern formation, we first consider the nonlinear refractive index induced in the atoms by the applied counterpropagating optical fields, \textit{i.e.}, below the threshold for pattern formation. In the next section, we extend this formalism to the above-threshold case to describe pattern formation in cold atoms.

A standard method for describing light-atom interactions is to define the material polarization, or the dipole moment per unit volume, of the atoms and to then solve the wave equation for the optical fields interacting with them. We consider a simplified $J=1/2\rightarrow J^\prime=3/2$ spin model for $^{87}$Rb. While we cannot realistically ignore the hyperfine structure, a fine-structure model is known to provide a good qualitative understanding of the experiment when most (${\sim}90\%$) of the atoms are tightly bunched and only undergo stretched-state transitions, as they are in our experiment~\cite{GreenbergOptExp}. We find that applying external magnetic fields to oppose background fields do not affect the threshold nor temporal persistence of pattern formation.

The material polarization for a gas of multi-level atoms with ground states $g\equiv\pm1/2$ and excited states $e\equiv\pm1/2,\pm3/2$ is
\begin{multline}
\vec{P}(z)=\sum_{g,e}\frac{\vec{\mu}_{ge}(\vec{\mu}_{eg}\cdot\vec{E})}{\hbar(\Delta+i/T_2)}\eta(z)\cdot\\
\frac{(\rho_{ee}-\rho_{gg})\left[1+(\omega-\omega_{eg})^2T_2^2\right]}{1+(\omega-\omega_{eg})^2T_2^2+(4/\hbar^2)|\vec{\mu}_{eg}\cdot \vec{E}|^2T_1T_2},
\end{multline}
where $\vec{\mu}_{ge}$ is the dipole moment, $\eta(z)$ is the density distribution, $\rho_{nm}$ are the density matrix elements, $T_1$ is the lifetime of the excited state, and $T_2$ is the characteristic dephasing time~\cite{Boyd}. We assume collisional dephasing is negligible and take $T_2=2T_1=1/\Gamma$, and we take all the population to be evenly distributed between the two ground states. We work in the regime where the intensity is much less than the off-resonant saturation intensity $I_{s\Delta}=I_{\text{sat}}(1+4\Delta^2\Gamma^2)$ so that the polarization becomes
\begin{multline}
\vec{P}(z)=\sum_{g,e}\Bigg\{-\frac{\vec{\mu}_{ge}(\vec{\mu}_{eg}\cdot\vec{E})(\Delta-i\Gamma/2)}{\hbar(\Delta^2+\Gamma^2/4)}\eta(z)\cdot\\\left[1-\frac{8}{\hbar^2\Gamma^2}\frac{|\vec{\mu}_{eg}\cdot \vec{E}|^2}{\left(1+4\Delta^2/\Gamma^2\right)}\right]\Bigg\}.
\label{polarizationmultilevel}
\end{multline}
The last term in square brackets represents the saturable nonlinearity. The density distribution $\eta(z)$ contains intensity-dependent terms that give rise to a bunching-induced nonlinearity. In the remainder of this section, we investigate these nonlinearities in the multi-level-atom picture to define the nonlinear refractive index for a gas of sub-Doppler-cooled atoms in a 1D optical lattice. We also show how this differs from the two-level-atom, lin$||$lin configuration described in Ref.~\cite{PhysRevA.90.013813}.

This model ignores Raman transitions between the $m_J=\pm1/2$ ground states, which is a good approximation when the atoms are tightly bunched at regions of $\hat{\sigma}^\pm$ polarization. However, this is not a good approximation for other experiments that use higher-temperature atoms~\cite{Labeyrie}, and any extension of this model to describe those systems should account for these transitions.

One of the main differences between the cold-atom case and the well-studied warm-atom case~\cite{Firth90} is the spatial dependence of $\eta(z)$. In warm atoms $\eta(z)\equiv n_a$, where $n_a$ is the average atomic density. In cold atoms, however, the atoms can spatially bunch in the dipole potential wells created by the optical fields, which results in a qualitatively different density distribution.

\begin{figure}
\begin{center}
\includegraphics[scale=0.3]{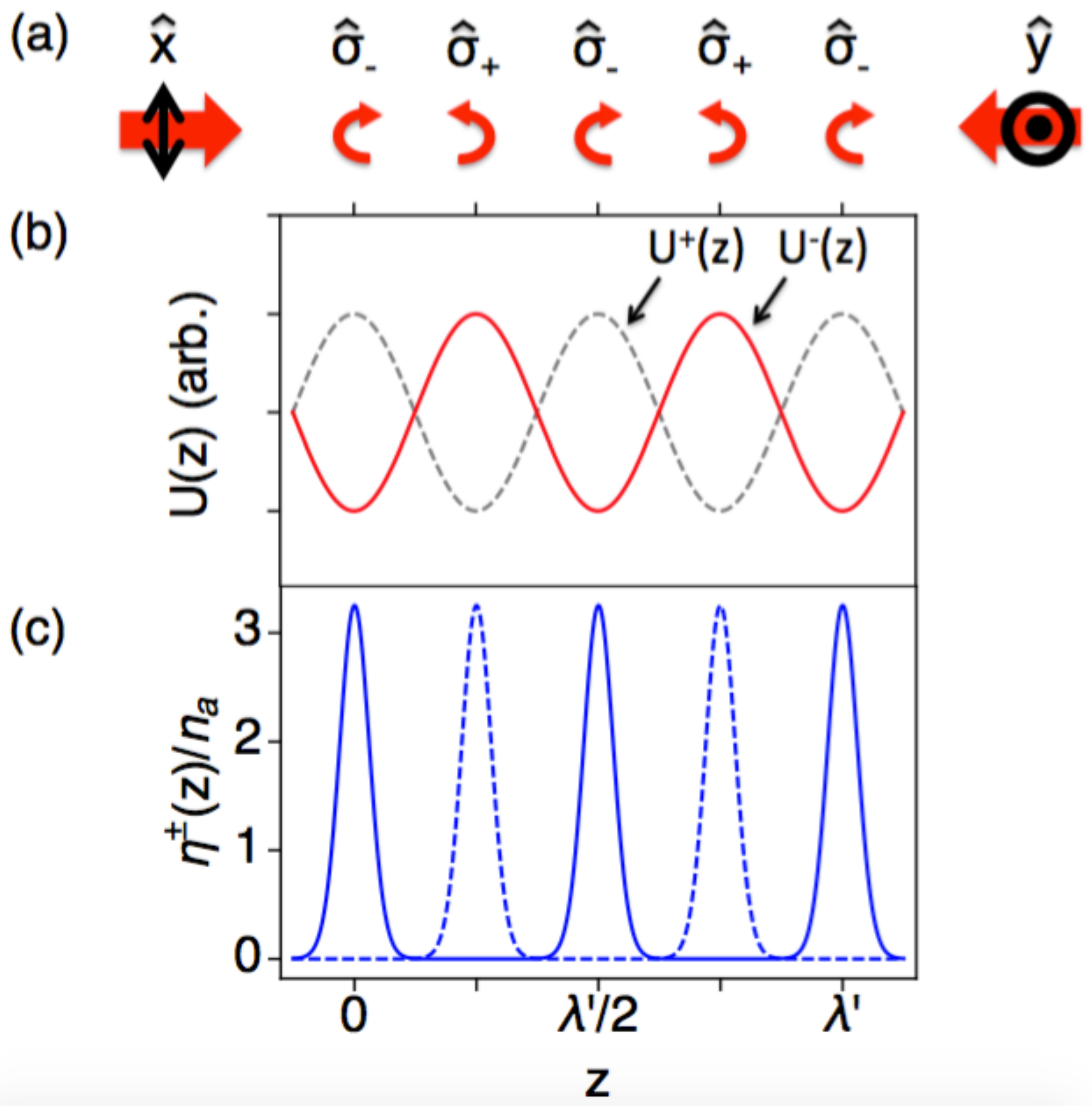}
\vspace{4mm}
\caption{\textbf{Dipole potential and density distributions with lin$\perp$lin electric field.} (a) The applied electric field polarization from Eq.~\ref{Efieldch7} periodically varies according to $\hat{\sigma}^-\rightarrow\hat{x}\rightarrow\hat{\sigma}^+\rightarrow\hat{y}\rightarrow\hat{\sigma}^-$. (b) The spatial variation of the dipole potentials $U^+(z)$ (gray, dashed) and $U^-(z)$ (red, solid) as a function of $z$, where $\lambda^\prime=2\pi/k^\prime$. (c) The density distributions $\eta^+(z)$ (solid) and $\eta^-(z)$ (dashed) normalized by $n_a$ for an example $\zeta=-7$.}
\label{dipoledensityfig}
\end{center}
\end{figure}

The density distribution $\eta(z)$ depends on the ratio of the dipole potential to the thermal energy of the atoms. The dipole potential $U(z)$ in the lin$\perp$lin polarization configuration is more complicated than in the lin$||$lin polarization configuration because the electric field polarization varies periodically in space. This periodically varying polarization gives rise to two superimposed light shifts
\begin{equation}
U^\pm(z)=\frac{4\Delta(\vec{\mu}^\pm_{eg}\cdot\vec{E})^*(\vec{\mu}^\pm_{eg}\cdot\vec{E})}{\hbar\Gamma^2(1+(2\Delta/\Gamma)^2)},
\label{dipolepotentiallinperplingen}
\end{equation}
which define the dipole potentials for atoms in the $m_J=\pm1/2$ ground states, respectively~\cite{Metcalf,GreenbergOptExp}, and where $\vec{\mu}^\pm_{eg}=\mu_{eg}\hat{\sigma}^{\pm}$. For atoms in the $m_J=+1/2$ ground state, there are two transitions that contribute to the dipole potential: the $m_J=+1/2\rightarrow m_{J^\prime}=+3/2$ transition (Clebsch-Gordon coefficient=1), and the $m_J=+1/2\rightarrow m_{J^\prime}=-1/2$ transition (Clebsch-Gordon coefficient$=\sqrt{1/3}$). We determine these separately according to $U^+(z)=U_{p,+1/2\rightarrow+3/2}^+(z)+U_{p,+1/2\rightarrow-1/2}^+(z)$. We note that $I_{sat}=\hbar^2\Gamma^2\epsilon_0c/2|\mu^\pm|^2$, where $\mu^\pm$ defines the dipole moment for a stretched-state transition. We define $F(z)=\tilde{F}e^{i\delta z}$ and $B(z)=\tilde{B}e^{-i\delta z}$, where $\delta$ defines the phase shift acquired by the field component as it propagates through the atoms. We define $I_F=2\epsilon_0c\tilde{F}^2$ and $I_B=2\epsilon_0c\tilde{B}^2$, where $c$ is the speed of light in vacuum and $\epsilon_0$ is the permittivity of free space, and we assume equal-intensity pump fields $I_F=I_B=I_p$. It follows that
\begin{equation}
U_{p,+1/2\rightarrow+3/2}^+(z)=\frac{\hbar\Delta I_p}{I_{\text{sat}}(1+(2\Delta/\Gamma)^2)}
-\frac{\hbar\Delta I_p\text{cos}(2k^\prime z)}{I_{\text{sat}}(1+(2\Delta/\Gamma)^2)}
\end{equation}
and
\begin{equation}
U_{p,+1/2\rightarrow-1/2}^+(z)=\frac{\hbar\Delta I_p}{3I_{\text{sat}}(1+(2\Delta/\Gamma)^2)}
+\frac{\hbar\Delta I_p\text{cos}(2k^\prime z)}{3I_{\text{sat}}(1+(2\Delta/\Gamma)^2)},
\end{equation}
where $k^\prime=k+\delta$ is the wavevector of the optical fields inside the medium. Here, $\delta=k\chi_{\text{eff}}/2$ is the phase shift acquired by the optical fields as they propagate through the atoms of index of refraction $n\simeq1+\chi_{\text{eff}}/2$, where $\chi_{\text{eff}}$ is the effective susceptibilty. The dipole potential for atoms in the $m_J=+1/2$ ground state is therefore
\begin{equation}
U_{p}^+(z)=\frac{4\hbar\Delta I_p}{3I_{\text{sat}}(1+(2\Delta/\Gamma)^2)}
-\frac{2\hbar\Delta I_p\text{cos}(2k^\prime z)}{3I_{\text{sat}}(1+(2\Delta/\Gamma)^2)}.
\end{equation}
We redefine this in terms of the total intensity $I_{\text{tot}}=2I_p$ and define
\begin{equation}
U_0=\frac{\hbar\Delta I_{\text{tot}}}{I_{\text{sat}}(1+(2\Delta/\Gamma)^2)},
\label{U0}
\end{equation}
so that the dipole potential can be rewritten as
\begin{equation}
U_{p}^+(z)=\frac{2}{3}U_0-\frac{1}{3}U_0\text{cos}(2k^\prime z),
\label{uplusinit}
\end{equation}
which agrees with the results of other sources that study Sisyphus cooling in a lin$\perp$lin polarization configuration~\cite{Dalibard,SisyphusCastinmain}. Here, only the second, spatially dependent term gives rise to a dipole force that contributes to atomic bunching. We introduce the variable $C^2=|C_{1/2,3/2}|^2-|C_{-1/2,1/2}|^2=|C_{-1/2,-3/2}|^2-|C_{1/2,-1/2}|^2=2/3$, which is the difference of the square of the Clebsch-Gordon coefficients for the two possible transitions, so that Eq.~\ref{uplusinit} and the analogous equation for the $m_J=-1/2$ ground state leads to the definitions
\begin{equation}
U^+(z)=U_0-C^2U_0\text{cos}^2(k^\prime z)
\label{Upluslinperplin}
\end{equation}
and
\begin{equation}
U^-(z)=U_0-C^2U_0\text{sin}^2(k^\prime z).
\label{Uminuslinperplin}
\end{equation}
These superimposed dipole potentials are phase-shifted by $\pi/2$ and their minima coincide with the locations of circular electric field polarizations depicted in Fig.~\ref{dipoledensityfig}(a). The phase-shifted potentials are depicted in Fig.~\ref{dipoledensityfig}(b), which show how each potential distribution varies with the electric field polarization. An example density distribution, which is derived in Sec.~\ref{bunchingnonlinmultilevel}, is given in Fig.~\ref{dipoledensityfig}(c) and shows how the atoms bunch at the locations of circular field polarizations when the dipole potential energy is greater than the thermal energy of the atoms.

\subsection{The bunching-induced nonlinearity}
\label{bunchingnonlinmultilevel}
We consider a steady-state, Maxwell-Boltzmann density distribution, which takes the general form $\eta(z)=\tilde{\eta}\text{exp}\left[-U(z)/k_BT\right]$. Based on the spatial periodicity of the superimposed dipole potentials defined in Eqs.~\ref{Upluslinperplin} and \ref{Uminuslinperplin}, we define the Floquet expansion
\begin{equation}
\eta(z)=\eta^+(z)+\eta^-(z)=\sum_{j=-\infty}^{\infty}\eta_j^+e^{2i(k^\prime z-\pi/2)*j}+\sum_{j=-\infty}^{\infty}\eta_j^-e^{2ik^\prime z*j},
\label{densitydistmultilevellinperplin}
\end{equation}
where
\begin{equation}
\eta_j^-=\frac{1}{\lambda^\prime/2}\int_{-\lambda^\prime/4}^{\lambda^\prime/4}\tilde{\eta}\text{exp}\left[\frac{-U^\pm(z)}{k_BT}\right]e^{-2ik^\prime z*j}dz.
\end{equation}
By defining $\eta_j^-$ this way, we take $\phi=-\pi/2$, which sets the electric field polarization to $\hat{\sigma}^-$ at $z=0$, and is used simply to impose a self-consistent density distribution. Also note that, according to the shift theorem, the coefficients $\eta_j^-=\eta_j^+$. Thus, $\eta_j^-$ becomes
\begin{equation}
\eta_j^-=\frac{1}{\lambda^\prime/2}\int_{-\lambda^\prime/4}^{\lambda^\prime/4}\tilde{\eta}\text{exp}\left[\frac{-2U_0/3-U_0\text{cos}(2k^\prime z)/3}{k_BT}\right]e^{-2ik^\prime z*j}dz.
\label{fcmid}
\end{equation}

We absorb the spatially independent part of the dipole potential into a new normalization constant $\eta^{\prime}=\tilde{\eta}\text{exp}\left[-2U_0/3k_BT\right]$. We define $\tilde{I}=I_p/I_{s\Delta}$, $\lambda^\prime=2\pi/k^\prime$, $\tilde{z}=2k^\prime z$, $d\tilde{z}=2k^\prime dz$, and the ratio
\begin{equation}
\zeta=\frac{C^2\tilde{\Delta}\tilde{I}}{\tilde{T}},
\label{zetamultilevel}
\end{equation}
where $\tilde{\Delta}=\Delta/\Gamma$, $\tilde{T}=T/T_D$, and $T_D=\hbar\Gamma/2k_B$. Equation~\ref{fcmid} then becomes
\begin{equation}
\eta_j^-=\frac{\eta^\prime}{2\pi}\int_{-\pi}^{\pi}\text{exp}\left[-\zeta\text{cos}(\tilde{z})\right]\text{cos}(\tilde{z}*j)d\tilde{z}.
\end{equation}
This yields
\begin{equation}
\eta_j^-=\eta^\prime I_j(-\zeta),
\end{equation}
where $I_j(x)$ are modified Bessel functions of the first kind of order $j$.

The normalization constant $\eta^\prime$ is calculated independently by equating the areal density to the integral of the density distribution over one period. In this case, unlike the $\text{lin}\big|\big|\text{lin}$ case in Ref.~\cite{PhysRevA.90.013813}, the areal density goes as $N_0=n_a*(\lambda^\prime/4)$, where $n_a$ is the average atomic density, because each pancake of atoms occurs every $\lambda^\prime/4$. Therefore,
\begin{equation}
\frac{n_a\lambda^\prime}{4}=\eta^\prime\int_{-\pi/2k^\prime}^{\pi/2k^\prime}\text{exp}\left[-\zeta\text{cos}(2k^\prime z)\right]dz.
\end{equation}
This yields
\begin{equation}
\eta^\prime=n_a/[2I_0(-\zeta)],
\label{normconstant}
\end{equation}
so the Fourier coefficients are given by
\begin{equation}
\eta_j^-=\frac{n_a}{2}\frac{I_j(-\zeta)}{I_0(-\zeta)}.
\label{fcs}
\end{equation}
The factor of $1/2$ here, which is different than the $\text{lin}\big|\big|\text{lin}$ case, corrects for the apparent double-counting of $\eta_0$ that occurs in our original definition of $\eta(z)$. Applying this normalization constant to Eq.~\ref{densitydistmultilevellinperplin}, the density distribution $\eta(z)$ is properly normalized for the lin$\perp$lin polarization configuration.

The density distribution in Eq.~\ref{densitydistmultilevellinperplin} contains highly nonlinear terms in the polarization of Eq.~\ref{polarizationmultilevel} for the tight-bunching regime; \textit{e.g.}, when $|\zeta|\gtrsim0.8$, the Bessel functions of Eq.~\ref{fcs} give rise to non-negligible $\chi^{(5)}$ and higher-order terms. Therefore, by working in the tight-bunching regime using sub-Doppler temperatures and small detunings, one can enhance the nonlinear index of refraction and study nonlinear optical effects even at low intensities.

To study the tight-bunching regime theoretically, it is necessary to use the normalization constant defined in Eq.~\ref{fcs}. This normalization constant provides self-consistency to our model, while other models that do not account for this normalization are restricted to the weak bunching regime~\cite{Muradyan}, where $|\zeta|\simeq1$. By properly normalizing the density distribution, one can predict the enhancement in the nonlinear index of refraction that is achievable by working in the tight-bunching regime, where $|\zeta|\gg1$. In order to solve for the explicit dependence of the index of refraction on $\zeta$, it is necessary to solve the wave equation for the coupled forward and backward field amplitudes $F(z)$ and $B(z)$.

\section{Calculating the refractive index for a lin$\perp$lin lattice}
To describe the light-atom interaction that gives rise to pattern formation, it is first necessary to define the index of refraction imposed in the atoms by the applied lin$\perp$lin optical lattice. The index of refraction is calculated by solving the wave equation
\begin{equation}
\nabla^2\overrightarrow{E}-\frac{1}{c^2}\frac{\partial^2\overrightarrow{E}}{\partial t^2}=\frac{1}{\epsilon_0 c^2}\frac{\partial^2\overrightarrow{P}}{\partial t^2}
\label{waveequation}
\end{equation}
for the forward and backward fields. We define
\begin{equation}
\vec{P}=\vec{P}^+_{g=+1/2}+\vec{P}^-_{g=-1/2}=\vec{P}^++\vec{P}^-
\label{polarizationseparationlatticelinperplin}
\end{equation}
and note that transitions from the two ground states decouple from one another, so that
\begin{multline}
P^+=\sum_{e}\Bigg\{-\frac{|\vec{\mu}_{g=+1/2,e}^\pm|(\vec{\mu}_{e,g=+1/2}^\pm\cdot\vec{E})(\Delta-i\Gamma/2)}{\hbar(\Delta^2+\Gamma^2/4)}\eta^\pm(z)\\\left[1-\frac{8}{\hbar^2\Gamma^2}\frac{|\vec{\mu}_{e,g=+1/2}^\pm\cdot \vec{E}|^2}{\left(1+4\Delta^2/\Gamma^2\right)}\right]\Bigg\},
\end{multline}
with an analogous expression for $P^-$. We then solve Eq.~\ref{waveequation} for $F(z)$ and $B(z)$ and find that the total phase shift acquired by each field is
\begin{equation}
\delta=\frac{k}{2}\chi_{\text{lin}}\Bigg[\left(1-3\tilde{I}\right)+\left(1-2\tilde{I}\right)\frac{I_1(-\zeta)}{I_0(-\zeta)}-2\tilde{I}\frac{I_2(-\zeta)}{I_0(-\zeta)}\Bigg],
\label{phaseshiftleftcirc}
\end{equation}
where $\chi_{\text{lin}}=-6\pi c^3(2\Delta/\Gamma)n_aC^2/[\omega_{eg}^3(1+4\Delta^2/\Gamma^2)]$ is the linear susceptibility. Therefore, the index of refraction in a lin$\perp$lin optical lattice goes as $n\simeq1+\chi_{\text{eff}}/2$ with
\begin{equation}
\chi_{\text{eff}}=\chi_{\text{lin}}\Bigg[\left(1-3\tilde{I}\right)+\left(1-2\tilde{I}\right)\frac{I_1(-\zeta)}{I_0(-\zeta)}-2\tilde{I}\frac{I_2(-\zeta)}{I_0(-\zeta)}\Bigg].
\label{chiefflinperplin}
\end{equation}

There are multiple qualitative differences between this case and the lin$||$lin case described in Ref.~\cite{PhysRevA.90.013813}. For blue detunings ($\Delta>0$) in the zero-temperature limit ($\zeta\rightarrow\infty$),
\begin{equation}
\text{lim}\underset{\zeta\rightarrow\infty}{\chi_{\text{eff}}}=-3\chi_{\text{lin}}\tilde{I}.
\label{deltalimitblue}
\end{equation}
Since we assume $\tilde{I}\ll1$, this implies that there is a finite but small light-atom interaction for blue detunings in the zero temperature limit. In contrast, in the lin$||$lin, two-level-atom case, this limit goes to zero because the atoms tightly bunch at the intensity zeroes. In contrast, the intensity is nonzero everywhere in the lin$\perp$lin polarization configuration. However, when we account for a multi-level atomic structure in a lin$\perp$lin configuration, where the intensity is always nonzero, the atoms always have a finite probability of interacting with the fields. For example, an atom in the $m_J=-1/2$ state minimizes its energy in the presence of a blue-detuned optical lattice by spatially bunching at a location of $\hat{\sigma}^+$-polarization. In this case, the atoms are not pumped into the stretched state. However, the atom can still be pumped via the weaker transition into the $m_{J^\prime}=+1/2$ excited state. Therefore, the conditions are never optimal for the atom to interact strongly with the optical fields, but the atoms may still always interact with them.

For red detunings ($\Delta<0$) in the zero temperature limit ($\zeta\rightarrow-\infty$),
\begin{equation}
\text{lim}\underset{\zeta\rightarrow-\infty}{\chi_{\text{eff}}}=\chi_{\text{lin}}\Bigg\{2-7\tilde{I}\Bigg\}.
\label{deltalimitred}
\end{equation}
The atoms thus experience a much larger phase shift for red detunings. This is the expected result because the atoms bunch into the regions of pure $\hat{\sigma}^\pm$ polarizations and continuously pump into the stretched states for red-detuned optical fields. In this case, unlike the blue-detuned case, the atoms are attracted to spatial locations where they interact strongly with the optical fields. It is for this reason that red detunings are optimal for achieving strong light-atom interactions in the regime of small detunings, sub-Doppler temperatures, and tight atomic bunching. We note that blue detunings are optimal in the low-intensity case with weakly bunched atoms~\cite{Muradyan}.

The index of refraction in the lin$\perp$lin polarization configuration is $n=n_{\text{lin}}+n_{\text{NL}}$, where $n_{\text{lin}}=1+\chi_{\text{lin}}C^2/2$ is the linear refractive index, and
\begin{equation}
n_{\text{NL}}=\frac{\chi_{\text{lin}}}{2}\Bigg[-3\tilde{I}+\left(1-2\tilde{I}\right)\frac{I_1(-\zeta)}{I_0(-\zeta)}-2\tilde{I}\frac{I_2(-\zeta)}{I_0(-\zeta)}\Bigg]
\label{chiefflinperplin}
\end{equation}
is the nonlinear refractive index. This represents a self-consistent definition of the index of refraction for a gas of sub-Doppler-cooled atoms in a lin$\perp$lin optical lattice, which can be used to predict and optimize the light-atom interaction strength\textemdash not limited to the zero-temperature limits discussed in this section.

Above a threshold value of $n_{\text{NL}}$, there exist transverse perturbations that give rise to transverse pattern formation. In order to define this threshold, we perform a stability analysis to derive the conditions under which transverse perturbations experience sufficient gain to form transverse patterns.

\section{Stability analysis}
A stability analysis is a standard technique for deriving the threshold condition for transverse optical pattern formation~\cite{Firth:88,0295-5075-18-8-005,PhysRevA.48.1610}. To simplify the stability analysis, we consider a two-spot optical pattern, whose beam geometry is depicted in Fig.~\ref{patterntheorypapersetupfig}(a), and whose electric field amplitude goes as
\begin{multline}
\overrightarrow{E}_0(z,r)={F}(z)e^{ikz}\hat{x}+e^{i\phi}{B}(z)e^{-ikz}\hat{y}+\\
e^{i\phi}\alpha{f}_+(z,r)e^{ik(\text{cos}\theta z-\text{sin}\theta r)}\hat{y}+e^{i\phi}\alpha{f}_-(z,r)e^{ik(\text{cos}\theta z+\text{sin}\theta r)}\hat{y}+\\
\alpha{b}_+(z,r)e^{ik(-\text{cos}\theta z-\text{sin}\theta r)}\hat{x}+\alpha{b}_-(z,r)e^{ik(-\text{cos}\theta z+\text{sin}\theta r)}\hat{x},
\label{Efieldstability}
\end{multline}
where we use $\alpha$ to keep track of the order of the small-amplitude generated fields. We use these polarizations because these are the relative polarizations we observe experimentally. However, it is straightforward to extend this analysis to study alternative polarization configurations. We again take $\phi=-\pi/2$ so that the polarization of all fields at $z=0$ is $\hat{\sigma}^-$-polarized.
With this modified electric field, the density distribution can be separated into those components of the dipole potential arising solely due to the pump-pump gratings and those arising due to the generated pump-pattern gratings. The density distribution defining atoms in the $m_J=\pm1/2$ ground states above threshold for pattern formation is
\begin{equation}
\overline{\eta}^\pm(z,r)=\eta^\pm(z)\text{exp}\left[-\frac{U_{\text{pp}}^\pm(z,r)}{k_BT}\right],
\end{equation}
where the argument of the second exponential contains terms of order $\alpha$ and defines the transverse density perturbations that arise due to the interference of a pump field and a pattern-forming field. We take the normalization constant to be that from Eq.~\ref{normconstant}, \textit{i.e.}, defined by the pump-pump gratings, which is a good approximation close to threshold.

To derive the threshold condition, we are interested in the regime where the generated fields are very weak. In this case, the pump-pattern dipole potentials are very shallow, so that the Taylor expansion
\begin{equation}
\text{exp}\left[-\frac{U_{\text{pp}}^\pm(z,r)}{k_BT}\right]\approx1-\frac{U_{\text{pp}}^\pm(z,r)}{k_BT}
\label{transversepert}
\end{equation}
is valid. Equation~\ref{transversepert} defines the transverse perturbations to the density distribution. Atoms that bunch into these perturbative gratings are termed ``self-organized'' because they are not imposed by externally applied optical fields~\cite{SchmittbergerMultimode}. These terms further enhance the material polarization and the power in the generated fields. In Eq.~\ref{transversepert},
\begin{equation}
U_{\text{pp}}^\pm(z,r)=\frac{4\Delta(\vec{\mu}^\pm_{eg}\cdot\vec{E})^*(\vec{\mu}^\pm_{eg}\cdot\vec{E})}{\hbar\Gamma^2(1+(2\Delta/\Gamma)^2)}\Bigg|_{O(\alpha)},
\end{equation}
but where we only retain those terms of order $\alpha$ that give rise to self-organized gratings. We neglect terms of order $\alpha^2$ or higher because those correspond to the interference of two generated fields, which are orders of magnitude weaker and negligible close to threshold. Simplifying, the polarization components become
\begin{multline}
\vec{P}^+(z,r)=-\frac{2\vec{\mu}_{ge}^+(\vec{\mu}_{eg}^+\cdot\vec{E})(2\tilde{\Delta}-i)}{\hbar\Gamma(1+4\tilde{\Delta}^2)}\left[1-\frac{8}{\hbar^2\Gamma^2}\frac{|\vec{\mu}_{eg}^+\cdot \vec{E}|^2}{(1+4\tilde{\Delta}^2)}\right]\cdot\\
\left[\sum_{j=-\infty}^{\infty}\eta_j^+e^{2i(k^\prime z-\pi/2)*j}\right]\left(1-\frac{8\tilde{\Delta}(\vec{\mu}^+_{eg}\cdot\vec{E}_{pp})^*(\vec{\mu}^+_{eg}\cdot\vec{E}_{pp})}{(\hbar\Gamma)^2\tilde{T}(1+4\tilde{\Delta}^2)}\right)
\label{Pfinit2}
\end{multline}
and
\begin{multline}
\vec{P}^-(z,r)=-\frac{2\vec{\mu}_{ge}^-(\vec{\mu}_{eg}^-\cdot\vec{E})(2\tilde{\Delta}-i)}{\hbar\Gamma(1+4\tilde{\Delta}^2)}\left[1-\frac{8}{\hbar^2\Gamma^2}\frac{|\vec{\mu}_{eg}^-\cdot \vec{E}|^2}{(1+4\tilde{\Delta}^2)}\right]\cdot\\\left[\sum_{j=-\infty}^{\infty}\eta_j^-e^{2ik^\prime z*j}\right]\left(1-\frac{8\tilde{\Delta}(\vec{\mu}^-_{eg}\cdot\vec{E}_{pp})^*(\vec{\mu}^-_{eg}\cdot\vec{E}_{pp})}{(\hbar\Gamma)^2\tilde{T}(1+4\tilde{\Delta}^2)}\right),
\label{Pfinit1}
\end{multline}
where the component $(\vec{\mu}^\pm_{eg}\cdot\vec{E}_{pp})^*(\vec{\mu}^\pm_{eg}\cdot\vec{E}_{pp})$ must consist of exactly one pump field term and one weak field term. We expand Eqs.~\ref{Pfinit2} and \ref{Pfinit1} in order to solve the wave equation.

Under the rotating wave approximation, the left-hand-side of the wave equation in the steady-state regime for the $f_+(z,r)$ weak field component that is $\hat{\sigma}^-$-polarized is
\begin{equation}
2ik\text{cos}\theta\frac{\partial f_+}{\partial z}e^{i(k\text{cos}\theta z-k\text{sin}\theta r-\omega t)}\frac{\hat{\sigma}^-}{\sqrt{2}},
\end{equation}
where we take the optical field amplitude variation in $r$ to be very small, so that $\partial f_+/\partial r\rightarrow0$.

To simplify the right-hand-side of the wave equation, we define the new variables
\begin{equation}
f_+(z,r)=f_+^\prime(z,r) e^{i(k^\prime-k\text{cos}\theta)z},
\end{equation}
\begin{equation}
f_-(z,r)=f_-^\prime(z,r) e^{i(k^\prime-k\text{cos}\theta)z},
\end{equation}
\begin{equation}
b_+(z,r)=b_+^\prime(z,r) e^{-i(k^\prime-k\text{cos}\theta)z},
\end{equation}
and
\begin{equation}
b_-(z,r)=b_-^\prime(z,r) e^{-i(k^\prime-k\text{cos}\theta)z}.
\end{equation}
We also define the following quantities:
\begin{equation}
\xi_0=\frac{k\chi_{\text{lin}}}{2\text{cos}\theta},
\end{equation}
\begin{multline}
A=2\eta_0-\frac{\tilde{\Delta}I_p}{2\tilde{T}I_{s\Delta}}\frac{2}{3}\left(5\eta_0+4\eta_1\right)-\frac{I_p}{2I_{s\Delta}}\frac{4}{3}\left(5\eta_0+4\eta_1\right)+\\
\frac{\tilde{\Delta}}{\tilde{T}}\left(\frac{I_p}{2I_{s\Delta}}\right)^2\frac{2}{9}\left(42\eta_0+52\eta_1+14\eta_2\right),
\end{multline}
\begin{multline}
B=-\frac{\tilde{\Delta}I_p}{2\tilde{T}I_{s\Delta}}\frac{2}{3}\left(5\eta_0+4\eta_1\right)-\frac{I_p}{2I_{s\Delta}}\frac{2}{3}\left(5\eta_0+4\eta_1\right)+\\
\frac{\tilde{\Delta}}{\tilde{T}}\left(\frac{I_p}{2I_{s\Delta}}\right)^2\frac{2}{9}\left(42\eta_0+52\eta_1+14\eta_2\right),
\end{multline}
\begin{multline}
C=\eta_1-\frac{\tilde{\Delta}I_p}{2\tilde{T}I_{s\Delta}}\frac{1}{3}\left(5\eta_0+8\eta_1+5\eta_2\right)+\\
-\frac{I_p}{2I_{s\Delta}}\frac{2}{3}\left(5\eta_0+8\eta_1+5\eta_2\right)+\\
\frac{\tilde{\Delta}}{\tilde{T}}\left(\frac{I_p}{2I_{s\Delta}}\right)^2\frac{1}{9}\left(56\eta_0+91\eta_1+56\eta_2+13\eta_3\right),
\end{multline}
and
\begin{multline}
D=-\frac{\tilde{\Delta}I_p}{2\tilde{T}I_{s\Delta}}\frac{1}{3}\left(5\eta_0+8\eta_1+5\eta_2\right)+\\
-\frac{I_p}{2I_{s\Delta}}\frac{1}{3}\left(5\eta_0+8\eta_1+5\eta_2\right)+\\
\frac{\tilde{\Delta}}{\tilde{T}}\left(\frac{I_p}{2I_{s\Delta}}\right)^2\frac{1}{9}\left(56\eta_0+91\eta_1+56\eta_2+13\eta_3\right).
\end{multline}
We extract only those terms that are phase-matched or nearly phase-matched to the generated fields to simplify the right-hand-side of the wave equation,\textit{e.g.}, we retain only those terms that oscillate at or close to $e^{i(k^\prime-k\text{cos}\theta)z}$ to solve the wave equation for $f_+^\prime(z,r)$. The resulting coupled amplitude equations are as follows with $\delta_k=k^\prime-k\text{cos}\theta$ and $\delta_A=\xi_0A-\delta_k$:
 \begin{equation}
 \frac{\partial}{\partial z}
 \begin{pmatrix}
  f_+^\prime\\
  f_-^{\prime*}\\
  b_+^\prime\\
  b_-^{\prime*}\\
 \end{pmatrix}=
\begin{pmatrix}
  i\delta_A & i\xi_0D & i\xi_0C & i\xi_0B \\
  -i\xi_0D &  -i\delta_A & -i\xi_0B & -i\xi_0C \\
  -i\xi_0C & -i\xi_0B &  -i\delta_A & -i\xi_0D \\
  i\xi_0B & i\xi_0C & i\xi_0D &  i\delta_A \\
  \end{pmatrix}
 \begin{pmatrix}
  f_+^\prime\\
  f_-^{\prime*}\\
  b_+^\prime\\
  b_-^{\prime*}\\
 \end{pmatrix}.
 \label{spinFPeqbunching}
 \end{equation}
We solve this equation for the boundary conditions $f_+^\prime(-L/2)=f_-^{\prime*}(-L/2)=b_+^\prime(+L/2)=b_-^{\prime*}(+L/2)=0$ using the methods in Refs.~\cite{PhysRevLett.48.1541,Firth90}.

\begin{figure}
\begin{center}
\includegraphics[scale=0.38]{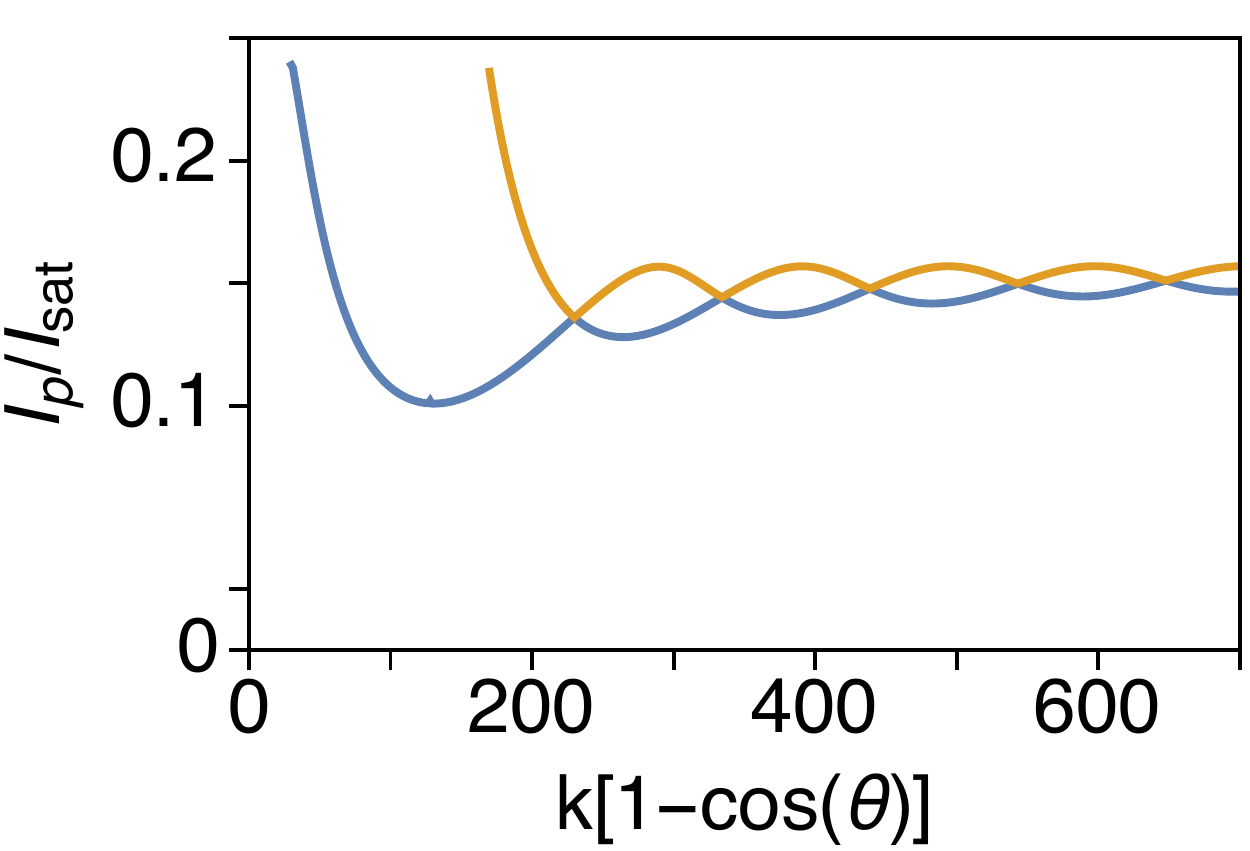}
\caption{\textbf{Predicted intensity vs. phase mismatch.} Single-beam intensity normalized by the resonant saturation intensity as a function of the phase mismatch $k(1-\text{cos}\theta)$, where again $k$ is the vacuum wavevector and $\theta$ is the angle between the generated fields and the applied (pump) fields. This is for the specific case of an optical depth of 20, $\tilde{\Delta}=-5$, $L=3$ cm, and $\tilde{T}=3/146$.}
\label{FirthPareMethodWithBunchingOD20Det5}
\end{center}
\end{figure}

For the example case of a detuning $\tilde{\Delta}=-5$, our experimentally measured temperature $T=3~\mu\text{K}$, and an optical depth ($\text{OD}$) of 20, where $\text{OD}=6\pi n_aL/k^2$ and we take $L=3~\text{cm}$, the solution is shown in Fig.~\ref{FirthPareMethodWithBunchingOD20Det5} for the predicted pump intensity as a function of the phase mismatch $k[1-\text{cos}(\theta)]$. The blue (lower) curve represents the minimum solution. The orange (higher) curve represents another higher numerical solution. This solution shows that, for $k[1-\text{cos}(\theta)]=0$ ($\theta=0$), the intensity required to generate new optical fields approaches infinity. However, at a finite angle, one can generate new optical fields. This is consistent with the concept of weak wave retardation discussed in Ref.~\cite{Chiao}, which tells us that the wave-mixing process that generates new fields requires a finite angle for phase-matching.

The minimum point on the blue curve in Fig.~\ref{FirthPareMethodWithBunchingOD20Det5} at $k[1-\text{cos}(\theta)]\simeq120$ provides a theoretical prediction for the angle of emission of the generated fields: $\theta\simeq5\text{ mrad}$, which is consistent with our typical experimental observations. The minimum point also predicts the minimum intensity threshold for these parameters: $I_p/I_{\text{sat}}\simeq0.1$. This predicted value is approximately an order of magnitude smaller than we observe experimentally, which we discuss below. 

We note that pattern formation only occurs for $n_{\text{NL}}>0$, \textit{i.e.}, a self-focusing nonlinearity~\cite{Chiao}. For warm atoms, the nonlinearity is self-focusing (defocusing) for blue (red) detunings, and thus theoretical models describing pattern formation in warm atoms consider only blue-detuned optical fields~\cite{Firth90}. However, in the low-intensity regime for cold atoms, the bunching-induced nonlinearity has a different detuning dependence and is self-focusing for red-detuned optical fields~\cite{PhysRevA.90.013813}.

\begin{figure}
\begin{center}
\includegraphics[scale=0.22]{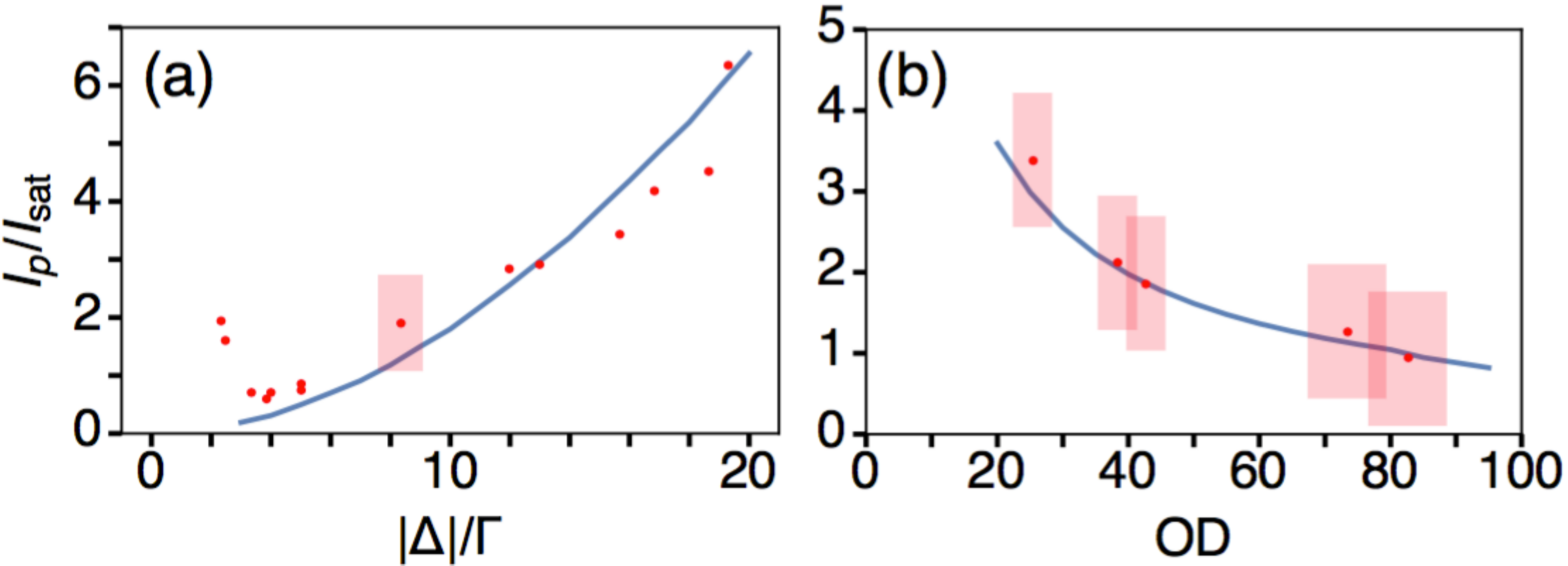}
\caption{\textbf{Results of stability analysis.} Single-beam intensity normalized by the resonant saturation intensity as a function of (a) the detuning normalized by the natural linewidth for the specific case of an optical depth of 62, $L=3$ cm, $\tilde{T}=3/146$, and red detunings with a free parameter value $p=4.5$, and (b) the optical depth for the specific case of an optical depth of $\tilde{\Delta}=-5$, $L=3$ cm, $\tilde{T}=3/146$, and a free parameter value $p=12$. The blue curve represents the predictions from Eq.~\ref{spinFPeqbunching} with free parameter $\tilde{I}\rightarrow p\tilde{I}$, and the red points represent experimental data. The red rectangle(s) represent the statistical error due to the intensity measurement with beam reshaping, the detuning measurement, and the confidence interval of the fit to the OD measurement.}
\label{threshvdetandOD}
\end{center}
\end{figure}

We solve Eq.~\ref{spinFPeqbunching} for other detunings and optical depths to build theoretical curves from the extracted solutions for the minimum predicted intensities. We show these theoretical results (blue) with experimental data points (red) in Fig.~\ref{threshvdetandOD} as functions of detuning and optical depth. To determine the experimental intensity threshold, we measure the power in the optical patterns and reduce the total pump intensity while keeping balanced intensities in the applied counterpropagating fields. We define the threshold intensity when we detect no power in the generated fields.

In the theoretical curves of Fig.~\ref{threshvdetandOD}, we incorporate a free parameter $p$ used to adjust the effective intensity $\tilde{I}\rightarrow p\tilde{I}$. The use of a free parameter adjusts the scale of the predicted curves in order to test whether the predicted curvature matches the experimental data. The deviation of the experiment from the scale of the theoretical predictions may arise due to multiple factors: 1. There are statistical errors in the initial measurements of $\tilde{I}$, $\tilde{\Delta}$, $L$, and the $\text{OD}$. 2. The beam reshaping effect discussed in the caption of Fig.~\ref{patterntheorypapersetupfig} introduces additional errors in $\tilde{I}$ and creates a nonuniform intensity throughout the atomic cloud, which is not treated in this plane wave model. 3. There are additional systematic errors in $L$ because the generated fields may emerge from the cloud before propagating its full length, thus reducing the effective length of the cloud. 4. This model assumes perfectly counterpropagating pump beams, and thus any slight mode-mismatch of the transverse pump profile reduces the efficiency of the wave-mixing process and increases the intensity threshold. 5. This model assumes a uniform atomic density across the pump beams, which is not the case when the pump beam size is comparable to the width of the atomic cloud, and 6. We do not explicitly account for Sisyphus cooling. For the data shown in Figs.~\ref{threshvdetandOD}(a) and (b), the best fits of the free parameter are $p\sim4.5$~and~$12$, respectively. The variation in the best fit of the free parameter between the two data sets arises because the sources of error discussed above (namely 2, 3, and 4) introduce systematic errors that vary day-to-day. With the use of a free parameter, the predicted theoretical curves fit well to our experimental data in Fig.~\ref{threshvdetandOD}. The observed increase in threshold for small detunings in Fig.~\ref{threshvdetandOD}(a) is attributed to increased absorption.

The observed intensity thresholds for pattern formation shown in Fig.~\ref{threshvdetandOD} are more than two orders of magnitude smaller than that observed in warm atoms~\cite{Dawes29042005} and more than one order of magnitude smaller than that observed in cold-atom experiments that use higher, but still sub-Doppler, temperatures~\cite{Labeyrie}. Our observation of ultra-low threshold powers is a manifestation of the enhancement of $n_{\text{NL}}$ that is achievable by using small detunings and far-sub-Doppler-cooled atoms. Our self-consistent model provides the means by which to study nonlinear optical effects in ultracold atoms, such as coupled optical/atomic pattern formation, where one can achieve enhanced light-atom interaction strengths and multimode self-organization.

\section{Conclusions}
In this article, we provide an overview of our experimental observation of pattern formation in cold atoms, and we describe the characteristics of the coupled optical/atomic patterns. We present a theoretical description for multi-level atoms in an optical lattice. We derive the index of refraction for sub-Doppler-cooled atoms in a lin$\perp$lin optical lattice, and we discuss how this differs from the lin$||$lin case. We then extend this model to a two-dimensional geometry with multiple optical fields in order to describe two-spot optical pattern formation. We then perform a stability analysis in order to derive the threshold condition for generating patterns, and we compare this to our experimental results.

This work represents the first stability analysis for pattern formation in cold atoms that allows for tight atomic bunching. This self-consistent description is useful for describing pattern formation experiments with strong light-atom interactions, in which one can study low-light-level nonlinear optics in cold atoms. This model also shows the importance of accounting for transverse perturbations in the density distribution, \textit{i.e.}, the self-organized atomic structures, which enhances the refractive index of the atoms during and above threshold for pattern formation and give rise to multimode atomic self-organization at low light levels.

\section{Funding Information}

We gratefully acknowledge the financial support of the National Science Foundation through Grant $\#$PHY-1206040.

\bibliography{patternstheorypaperreferences}

\begin{thebibliography}{27}
\expandafter\ifx\csname natexlab\endcsname\relax\def\natexlab#1{#1}\fi
\expandafter\ifx\csname bibnamefont\endcsname\relax
  \def\bibnamefont#1{#1}\fi
\expandafter\ifx\csname bibfnamefont\endcsname\relax
  \def\bibfnamefont#1{#1}\fi
\expandafter\ifx\csname citenamefont\endcsname\relax
  \def\citenamefont#1{#1}\fi
\expandafter\ifx\csname url\endcsname\relax
  \def\url#1{\texttt{#1}}\fi
\expandafter\ifx\csname urlprefix\endcsname\relax\def\urlprefix{URL }\fi
\providecommand{\bibinfo}[2]{#2}
\providecommand{\eprint}[2][]{\url{#2}}

\bibitem[{\citenamefont{Dawes et~al.}(2010)\citenamefont{Dawes, Gauthier,
  Schumacher, Kwong, Binder, and Smirl}}]{DawesReview}
\bibinfo{author}{\bibfnamefont{A.~M.~C.} \bibnamefont{Dawes}},
  \bibinfo{author}{\bibfnamefont{D.~J.} \bibnamefont{Gauthier}},
  \bibinfo{author}{\bibfnamefont{S.}~\bibnamefont{Schumacher}},
  \bibinfo{author}{\bibfnamefont{N.~H.} \bibnamefont{Kwong}},
  \bibinfo{author}{\bibfnamefont{R.}~\bibnamefont{Binder}}, \bibnamefont{and}
  \bibinfo{author}{\bibfnamefont{A.~L.} \bibnamefont{Smirl}},
  \bibinfo{journal}{Laser \& Photonics Reviews} \textbf{\bibinfo{volume}{4}},
  \bibinfo{pages}{221} (\bibinfo{year}{2010}), ISSN \bibinfo{issn}{1863-8899}.

\bibitem[{\citenamefont{Greenberg et~al.}(2011)\citenamefont{Greenberg,
  Schmittberger, and Gauthier}}]{GreenbergOptExp}
\bibinfo{author}{\bibfnamefont{J.~A.} \bibnamefont{Greenberg}},
  \bibinfo{author}{\bibfnamefont{B.~L.} \bibnamefont{Schmittberger}},
  \bibnamefont{and} \bibinfo{author}{\bibfnamefont{D.~J.}
  \bibnamefont{Gauthier}}, \bibinfo{journal}{Opt. Express}
  \textbf{\bibinfo{volume}{19}}, \bibinfo{pages}{22535} (\bibinfo{year}{2011}).

\bibitem[{\citenamefont{Schilke et~al.}(2012)\citenamefont{Schilke, Zimmerman,
  Courteille, and Guerin}}]{SchilkeExpTransverse}
\bibinfo{author}{\bibfnamefont{A.}~\bibnamefont{Schilke}},
  \bibinfo{author}{\bibfnamefont{C.}~\bibnamefont{Zimmerman}},
  \bibinfo{author}{\bibfnamefont{P.~W.} \bibnamefont{Courteille}},
  \bibnamefont{and} \bibinfo{author}{\bibfnamefont{W.}~\bibnamefont{Guerin}},
  \bibinfo{journal}{Nat. Photon.} \textbf{\bibinfo{volume}{6}},
  \bibinfo{pages}{101} (\bibinfo{year}{2012}).

\bibitem[{\citenamefont{Labeyrie et~al.}(2014)\citenamefont{Labeyrie, Tesio,
  Gomes, Oppo, Firth, Robb, Arnold, Kaiser, and Ackemann}}]{Labeyrie}
\bibinfo{author}{\bibfnamefont{G.}~\bibnamefont{Labeyrie}},
  \bibinfo{author}{\bibfnamefont{E.}~\bibnamefont{Tesio}},
  \bibinfo{author}{\bibfnamefont{P.~M.} \bibnamefont{Gomes}},
  \bibinfo{author}{\bibfnamefont{G.-L.} \bibnamefont{Oppo}},
  \bibinfo{author}{\bibfnamefont{W.~J.} \bibnamefont{Firth}},
  \bibinfo{author}{\bibfnamefont{G.~R.~M.} \bibnamefont{Robb}},
  \bibinfo{author}{\bibfnamefont{A.~S.} \bibnamefont{Arnold}},
  \bibinfo{author}{\bibfnamefont{R.}~\bibnamefont{Kaiser}}, \bibnamefont{and}
  \bibinfo{author}{\bibfnamefont{T.}~\bibnamefont{Ackemann}},
  \bibinfo{journal}{Nat. Photon.} \textbf{\bibinfo{volume}{8}},
  \bibinfo{pages}{321} (\bibinfo{year}{2014}).

\bibitem[{\citenamefont{Grynberg}(1988)}]{Grynberg1988321}
\bibinfo{author}{\bibfnamefont{G.}~\bibnamefont{Grynberg}},
  \bibinfo{journal}{Optics Communications} \textbf{\bibinfo{volume}{66}},
  \bibinfo{pages}{321 } (\bibinfo{year}{1988}), ISSN \bibinfo{issn}{0030-4018}.

\bibitem[{\citenamefont{Firth et~al.}(1990)\citenamefont{Firth, Par\'{e}, and
  Fitzgerald}}]{Firth90}
\bibinfo{author}{\bibfnamefont{W.~J.} \bibnamefont{Firth}},
  \bibinfo{author}{\bibfnamefont{C.}~\bibnamefont{Par\'{e}}}, \bibnamefont{and}
  \bibinfo{author}{\bibfnamefont{A.}~\bibnamefont{Fitzgerald}},
  \bibinfo{journal}{J. Opt. Soc. Am. B} \textbf{\bibinfo{volume}{7}},
  \bibinfo{pages}{1087} (\bibinfo{year}{1990}).

\bibitem[{\citenamefont{Dawes et~al.}(2005)\citenamefont{Dawes, Illing, Clark,
  and Gauthier}}]{Dawes29042005}
\bibinfo{author}{\bibfnamefont{A.~M.~C.} \bibnamefont{Dawes}},
  \bibinfo{author}{\bibfnamefont{L.}~\bibnamefont{Illing}},
  \bibinfo{author}{\bibfnamefont{S.~M.} \bibnamefont{Clark}}, \bibnamefont{and}
  \bibinfo{author}{\bibfnamefont{D.~J.} \bibnamefont{Gauthier}},
  \bibinfo{journal}{Science} \textbf{\bibinfo{volume}{308}},
  \bibinfo{pages}{672} (\bibinfo{year}{2005}).

\bibitem[{\citenamefont{Black et~al.}(2003)\citenamefont{Black, Chan, and
  Vuleti\ifmmode~\acute{c}\else \'{c}\fi{}}}]{PhysRevLett.91.203001}
\bibinfo{author}{\bibfnamefont{A.~T.} \bibnamefont{Black}},
  \bibinfo{author}{\bibfnamefont{H.~W.} \bibnamefont{Chan}}, \bibnamefont{and}
  \bibinfo{author}{\bibfnamefont{V.}~\bibnamefont{Vuleti\ifmmode~\acute{c}\else
  \'{c}\fi{}}}, \bibinfo{journal}{Phys. Rev. Lett.}
  \textbf{\bibinfo{volume}{91}}, \bibinfo{pages}{203001}
  (\bibinfo{year}{2003}).

\bibitem[{\citenamefont{Baumann et~al.}(2010)\citenamefont{Baumann, Guerlin,
  Brennecke, and Esslinger}}]{BaumannSelfOrganization}
\bibinfo{author}{\bibfnamefont{K.}~\bibnamefont{Baumann}},
  \bibinfo{author}{\bibfnamefont{C.}~\bibnamefont{Guerlin}},
  \bibinfo{author}{\bibfnamefont{F.}~\bibnamefont{Brennecke}},
  \bibnamefont{and}
  \bibinfo{author}{\bibfnamefont{T.}~\bibnamefont{Esslinger}},
  \bibinfo{journal}{Nature} \textbf{\bibinfo{volume}{464}},
  \bibinfo{pages}{1301} (\bibinfo{year}{2010}).

\bibitem[{\citenamefont{Gopalakrishnan
  et~al.}(2009)\citenamefont{Gopalakrishnan, Lev, and
  Goldbart}}]{GopLevGoldNatPhys}
\bibinfo{author}{\bibfnamefont{S.}~\bibnamefont{Gopalakrishnan}},
  \bibinfo{author}{\bibfnamefont{B.~L.} \bibnamefont{Lev}}, \bibnamefont{and}
  \bibinfo{author}{\bibfnamefont{P.~M.} \bibnamefont{Goldbart}},
  \bibinfo{journal}{Nat. Phys.} \textbf{\bibinfo{volume}{5}},
  \bibinfo{pages}{845} (\bibinfo{year}{2009}).

\bibitem[{\citenamefont{Schmittberger and
  Gauthier}(2016)}]{SchmittbergerMultimode}
\bibinfo{author}{\bibfnamefont{B.~L.} \bibnamefont{Schmittberger}}
  \bibnamefont{and} \bibinfo{author}{\bibfnamefont{D.~J.}
  \bibnamefont{Gauthier}}, \bibinfo{journal}{arXiv:1603.06294}
  (\bibinfo{year}{2016}).

\bibitem[{\citenamefont{Muradyan et~al.}(2005)\citenamefont{Muradyan, Wang,
  Williams, and Saffman}}]{Muradyan}
\bibinfo{author}{\bibfnamefont{G.~A.} \bibnamefont{Muradyan}},
  \bibinfo{author}{\bibfnamefont{Y.}~\bibnamefont{Wang}},
  \bibinfo{author}{\bibfnamefont{W.}~\bibnamefont{Williams}}, \bibnamefont{and}
  \bibinfo{author}{\bibfnamefont{M.}~\bibnamefont{Saffman}}, in
  \emph{\bibinfo{booktitle}{Nonlinear Guided Waves and Their Applications}}
  (\bibinfo{publisher}{Optical Society of America}, \bibinfo{year}{2005}), p.
  \bibinfo{pages}{ThB29}.

\bibitem[{\citenamefont{Greenberg and Gauthier}(2012)}]{GreenbergEPL}
\bibinfo{author}{\bibfnamefont{J.~A.} \bibnamefont{Greenberg}}
  \bibnamefont{and} \bibinfo{author}{\bibfnamefont{D.~J.}
  \bibnamefont{Gauthier}}, \bibinfo{journal}{EPL (Europhysics Letters)}
  \textbf{\bibinfo{volume}{98}}, \bibinfo{pages}{24001} (\bibinfo{year}{2012}),
  \urlprefix\url{http://stacks.iop.org/0295-5075/98/i=2/a=24001}.

\bibitem[{\citenamefont{Tesio et~al.}(2014)\citenamefont{Tesio, Robb, Ackemann,
  Firth, and Oppo}}]{PhysRevLett.112.043901}
\bibinfo{author}{\bibfnamefont{E.}~\bibnamefont{Tesio}},
  \bibinfo{author}{\bibfnamefont{G.~R.~M.} \bibnamefont{Robb}},
  \bibinfo{author}{\bibfnamefont{T.}~\bibnamefont{Ackemann}},
  \bibinfo{author}{\bibfnamefont{W.~J.} \bibnamefont{Firth}}, \bibnamefont{and}
  \bibinfo{author}{\bibfnamefont{G.-L.} \bibnamefont{Oppo}},
  \bibinfo{journal}{Phys. Rev. Lett.} \textbf{\bibinfo{volume}{112}},
  \bibinfo{pages}{043901} (\bibinfo{year}{2014}),
  \urlprefix\url{http://link.aps.org/doi/10.1103/PhysRevLett.112.043901}.

\bibitem[{\citenamefont{Schmittberger and Gauthier}(2014)}]{PhysRevA.90.013813}
\bibinfo{author}{\bibfnamefont{B.~L.} \bibnamefont{Schmittberger}}
  \bibnamefont{and} \bibinfo{author}{\bibfnamefont{D.~J.}
  \bibnamefont{Gauthier}}, \bibinfo{journal}{Phys. Rev. A}
  \textbf{\bibinfo{volume}{90}}, \bibinfo{pages}{013813}
  (\bibinfo{year}{2014}).

\bibitem[{\citenamefont{Greenberg et~al.}(2007)\citenamefont{Greenberg, Oria,
  Dawes, and Gauthier}}]{Greenberg:07}
\bibinfo{author}{\bibfnamefont{J.~A.} \bibnamefont{Greenberg}},
  \bibinfo{author}{\bibfnamefont{M.}~\bibnamefont{Oria}},
  \bibinfo{author}{\bibfnamefont{A.~M.~C.} \bibnamefont{Dawes}},
  \bibnamefont{and} \bibinfo{author}{\bibfnamefont{D.~J.}
  \bibnamefont{Gauthier}}, \bibinfo{journal}{Opt. Express}
  \textbf{\bibinfo{volume}{15}}, \bibinfo{pages}{17699} (\bibinfo{year}{2007}).

\bibitem[{\citenamefont{Yariv and Pepper}(1977)}]{Yariv:77}
\bibinfo{author}{\bibfnamefont{A.}~\bibnamefont{Yariv}} \bibnamefont{and}
  \bibinfo{author}{\bibfnamefont{D.~M.} \bibnamefont{Pepper}},
  \bibinfo{journal}{Opt. Lett.} \textbf{\bibinfo{volume}{1}},
  \bibinfo{pages}{16} (\bibinfo{year}{1977}).

\bibitem[{\citenamefont{Chiao et~al.}(1966)\citenamefont{Chiao, Kelley, and
  Garmire}}]{Chiao}
\bibinfo{author}{\bibfnamefont{R.~Y.} \bibnamefont{Chiao}},
  \bibinfo{author}{\bibfnamefont{P.~L.} \bibnamefont{Kelley}},
  \bibnamefont{and} \bibinfo{author}{\bibfnamefont{E.}~\bibnamefont{Garmire}},
  \bibinfo{journal}{Phys. Rev. Lett.} \textbf{\bibinfo{volume}{17}},
  \bibinfo{pages}{1158} (\bibinfo{year}{1966}).

\bibitem[{\citenamefont{Jersblad et~al.}(2004)\citenamefont{Jersblad, Ellmann,
  St\o{}chkel, Kastberg, Sanchez-Palencia, and Kaiser}}]{PhysRevA.69.013410}
\bibinfo{author}{\bibfnamefont{J.}~\bibnamefont{Jersblad}},
  \bibinfo{author}{\bibfnamefont{H.}~\bibnamefont{Ellmann}},
  \bibinfo{author}{\bibfnamefont{K.}~\bibnamefont{St\o{}chkel}},
  \bibinfo{author}{\bibfnamefont{A.}~\bibnamefont{Kastberg}},
  \bibinfo{author}{\bibfnamefont{L.}~\bibnamefont{Sanchez-Palencia}},
  \bibnamefont{and} \bibinfo{author}{\bibfnamefont{R.}~\bibnamefont{Kaiser}},
  \bibinfo{journal}{Phys. Rev. A} \textbf{\bibinfo{volume}{69}},
  \bibinfo{pages}{013410} (\bibinfo{year}{2004}).

\bibitem[{\citenamefont{Boyd}(2008)}]{Boyd}
\bibinfo{author}{\bibfnamefont{R.~W.} \bibnamefont{Boyd}},
  \emph{\bibinfo{title}{Nonlinear Optics, 3rd Ed.}}
  (\bibinfo{publisher}{Academic Press}, \bibinfo{year}{2008}).

\bibitem[{\citenamefont{Metcalf and van~der Straten}(1999)}]{Metcalf}
\bibinfo{author}{\bibfnamefont{H.~J.} \bibnamefont{Metcalf}} \bibnamefont{and}
  \bibinfo{author}{\bibfnamefont{P.}~\bibnamefont{van~der Straten}},
  \emph{\bibinfo{title}{Laser Cooling and Trapping}}
  (\bibinfo{publisher}{Springer}, \bibinfo{year}{1999}).

\bibitem[{\citenamefont{Dalibard and Cohen-Tannoudji}(1989)}]{Dalibard}
\bibinfo{author}{\bibfnamefont{J.}~\bibnamefont{Dalibard}} \bibnamefont{and}
  \bibinfo{author}{\bibfnamefont{C.}~\bibnamefont{Cohen-Tannoudji}},
  \bibinfo{journal}{J. Opt. Soc. Am. B} \textbf{\bibinfo{volume}{6}},
  \bibinfo{pages}{2023} (\bibinfo{year}{1989}).

\bibitem[{\citenamefont{Castin et~al.}(1991)\citenamefont{Castin, Dalibard, and
  Cohen-Tannoudji}}]{SisyphusCastinmain}
\bibinfo{author}{\bibfnamefont{Y.}~\bibnamefont{Castin}},
  \bibinfo{author}{\bibfnamefont{J.}~\bibnamefont{Dalibard}}, \bibnamefont{and}
  \bibinfo{author}{\bibfnamefont{C.}~\bibnamefont{Cohen-Tannoudji}},
  \bibinfo{journal}{in Light Induced Kinetic Effects in Atoms, Ions and
  Molecules, Eds. L. Moi \textit{et al.}, (ETS Editrice, Pisa, Italy)}
  (\bibinfo{year}{1991}).

\bibitem[{\citenamefont{Firth and Par\'{e}}(1988)}]{Firth:88}
\bibinfo{author}{\bibfnamefont{W.~J.} \bibnamefont{Firth}} \bibnamefont{and}
  \bibinfo{author}{\bibfnamefont{C.}~\bibnamefont{Par\'{e}}},
  \bibinfo{journal}{Opt. Lett.} \textbf{\bibinfo{volume}{13}},
  \bibinfo{pages}{1096} (\bibinfo{year}{1988}).

\bibitem[{\citenamefont{Petrossian et~al.}(1992)\citenamefont{Petrossian,
  Pinard, Ma\^{i}tre, Courtois, and Grynberg}}]{0295-5075-18-8-005}
\bibinfo{author}{\bibfnamefont{A.}~\bibnamefont{Petrossian}},
  \bibinfo{author}{\bibfnamefont{M.}~\bibnamefont{Pinard}},
  \bibinfo{author}{\bibfnamefont{A.}~\bibnamefont{Ma\^{i}tre}},
  \bibinfo{author}{\bibfnamefont{J.-Y.} \bibnamefont{Courtois}},
  \bibnamefont{and} \bibinfo{author}{\bibfnamefont{G.}~\bibnamefont{Grynberg}},
  \bibinfo{journal}{EPL (Europhysics Letters)} \textbf{\bibinfo{volume}{18}},
  \bibinfo{pages}{689} (\bibinfo{year}{1992}).

\bibitem[{\citenamefont{Gaeta and Boyd}(1993)}]{PhysRevA.48.1610}
\bibinfo{author}{\bibfnamefont{A.~L.} \bibnamefont{Gaeta}} \bibnamefont{and}
  \bibinfo{author}{\bibfnamefont{R.~W.} \bibnamefont{Boyd}},
  \bibinfo{journal}{Phys. Rev. A} \textbf{\bibinfo{volume}{48}},
  \bibinfo{pages}{1610} (\bibinfo{year}{1993}).

\bibitem[{\citenamefont{Silberberg and Bar-Joseph}(1982)}]{PhysRevLett.48.1541}
\bibinfo{author}{\bibfnamefont{Y.}~\bibnamefont{Silberberg}} \bibnamefont{and}
  \bibinfo{author}{\bibfnamefont{I.}~\bibnamefont{Bar-Joseph}},
  \bibinfo{journal}{Phys. Rev. Lett.} \textbf{\bibinfo{volume}{48}},
  \bibinfo{pages}{1541} (\bibinfo{year}{1982}).

\end{thebibliography}

%
%

\end{document}